\newif\ifcomment

\commentfalse

\ifcomment
    \documentclass[useAMS,usenatbib,onecolumn]{mn2e}
    \usepackage{charter} % bigger and rounder
\else
    \documentclass[useAMS,usenatbib]{mn2e}
    \usepackage{times} % closer to actual MNRAS journal
\fi

\usepackage{graphicx}
\usepackage{calc}
\usepackage{amssymb}
\usepackage{amsmath}
\title[Galaxy Infall In Modified Gravity]{Galaxy Infall Kinematics as a Test of Modified Gravity}
\author[Zu et al.]{ 
\parbox{\textwidth}{
Ying Zu$^1$\thanks{E-mail: yingzu@astronomy.ohio-state.edu},
David H. Weinberg$^{1}$,
Elise Jennings$^{2}$,
Baojiu Li$^{3}$,
Mark Wyman$^{2,4}$,
}
\vspace*{4pt} \\
${1}$~Department of Astronomy and CCAPP, The Ohio State University, 140 W. 18th Avenue,
Columbus, OH 43210, USA\\
${2}$~The Kavli Institue for Cosmological Physics and Enrico Fermi Institute,
University of Chicago, 5640 South Ellis Avenue, Chicago, IL 60637, USA\\
${3}$~Institute for Computational Cosmology, Department of Physics, Durham
University, Durham DH1 3LE, UK\\
${4}$~Center for Cosmology and Particle Physics, Department of Physics,
    New York University, New York, NY, 10003, U.S.A.\\
}
\def\lcdm{\Lambda\mathrm{CDM}}

\def\xicgs{\xi^{s}_{cg}}
\def\xicgr{\xi^{r}_{cg}}
\def\mpc{\mathrm{Mpc}}
\def\hmpc{h^{-1}\mathrm{Mpc}}

\def\hkpc{h^{-1}\mathrm{kpc}}
\def\kpc{\mathrm{kpc}}
\def\hmsol{h^{-1}M_\odot}

\def\p2d{P(v_r, v_t)}

\def\vrc{v_{r, c}}
\def\f{f_{\mathrm{vir}}}

\def\sr{\sigma_{\mathrm{rad}}}
\def\st{\sigma_{\mathrm{tan}}}
\def\s0{\sigma_{0}}
\def\rpic{r_{\pi, c}}

\def\fr{f(R)}
\def\kms{\mathrm{km}/\mathrm{s}}
\def\f0{f_{R0}}
\def\om{\Omega_m}
\def\s8{\sigma_8}
\def\ol{\Omega_\Lambda}

\ifcomment
\fi

\newlength{\figtextwidth}
\begin{document} 
\ifcomment
    \fontsize{12pt}{12pt}\selectfont % set font size in comment mode
    \setlength{\figtextwidth}{\textwidth + 5.5cm}%
\else
    \setlength{\figtextwidth}{\textwidth + 0.0cm}%
\fi
\date{\today} \maketitle
%%%%%%%%%%%%%%%%%%%%%%%%%%%%%%%%%%%%%%%%%%%%%%%%%%%%%%%
%
%%%%%%%%%%%%%%%%%%%%%%%%%%%%%%%%%%%%%%%%%%%%%%%%%%%%%%%
%%  Abstract and Keywords
%%%%%%%%%%%%%%%%%%%%%%%%%%%%%%%%%%%%%%%%%%%%%%%%%%%%%%%
\begin{abstract}
    Infrared modifications of General Relativity~(GR) can be revealed by comparing the mass of galaxy clusters
    estimated from weak lensing to that from infall kinematics. We measure the 2D galaxy velocity distribution in
    the cluster infall region by applying the galaxy infall kinematics~(GIK)
    model developed by \cite{zu2013} to two
    suites of $\fr$ and Galileon modified gravity simulations. Despite having distinct screening mechanisms, namely,
    the Chameleon and the Vainshtein effects, the $\fr$ and Galileon clusters exhibit very similar deviations in
    their GIK profiles from GR, with $\sim 100\text{--}200\,\kms$ enhancement in the characteristic infall velocity at
    $r=5\,\hmpc$ and $50\text{--}100\,\kms$ broadening in the radial and tangential velocity dispersions across the
    entire infall region, for clusters with mass $\sim10^{14}\,\hmsol$ at $z=0.25$. These deviations are detectable
    via the GIK reconstruction of the redshift--space cluster--galaxy cross--correlation function, $\xicgs(r_p,r_\pi)$,
    which shows $\sim 1\text{--}2\,\hmpc$ increase in the characteristic line-of-sight distance $\rpic$ at
     $r_p<6\,\hmpc$ from GR predictions. With overlapping deep imaging and large redshift surveys in the
    future, we expect that the GIK modelling of $\xicgs$, in combination with the stacked weak lensing measurements,
    will provide powerful diagnostics of modified gravity theories and the origin of cosmic acceleration.
\end{abstract}
%%%%%%%%%%%%%%%%%%%%%%%%%%%%%%%%%%%%%%%%%%%%%%%%%%%%%%%
%%%%%%%%%%%%%%%%%%%%%%%%%%%%%%%%%%%%%%%%%%%%%%%%%%%%%%%
%%%%%%%%%%%%%%%%%%%%%%%%%%%%%%%%%%%%%%%%%%%%%%%%%%%%%%%

\section{Introduction}
\label{sec:intro}

The late--time acceleration of the Universe can be explained by modifying General Relativity~(GR) on cosmological
scales, avoiding the need of invoking a cosmological constant $\Lambda$ or an exotic repulsive fluid~(a.k.a.,
dark energy).  Many popular modified gravity~(MG) theories rely on an extra scalar field $\psi$ to mediate a fifth
force,\footnote{Theories such as the braneworld and massive/resonant gravity models can be reduced to scalar--tensor
theories in the ``decoupling limit''~\citep{luty2003, de_rham2011-1, hinterbichler2012}.} making the distributions
and motions of galaxies different from those predicted by GR~\citep[see][and references therein]{jain2010}. In
particular, the coherent infall of galaxies onto massive clusters will exhibit systematic deviations due to
the enhanced gravitational forces. \citeauthor{zu2013} (2013, hereafter ZW13) demonstrated that in $\lcdm$+GR
simulations the velocity distribution of galaxies in the virial and infall regions of clusters~(hereafter abbreviated
GIK for galaxy infall kinematics) is well described by a 2--component velocity distribution model, which can be
reconstructed from measurements of the redshift--space cluster--galaxy cross--correlation function, $\xicgs$. In
this paper, we apply GIK modelling to two suites of different MG simulations and investigate the possible signals
of MG imprinted on the redshift--space distribution of galaxies around clusters, using dark matter particles and
halos as proxies for galaxies. (For more general discussions of clusters as tests of cosmic acceleration theories
we refer readers to \cite{weinberg2013}, and for a succinct discussion of distinguishing MG from dark energy
to \citeauthor{hu2009} 2009.)

While deviations from GR may be welcomed on cosmological scales, a ``screening'' mechanism must be invoked in MG
theories to recover GR in high density regions like the solar system, where GR has passed numerous stringent
tests~\citep[e.g., ][]{bertotti2003, baesler1999}. Current viable screening mechanisms generally fall into two classes:
\begin{itemize}
\item The Chameleon--like mechanism, in which the self--interactions of
    the scalar field are regulated by a potential $V(\psi)$~\citep{buchdahl1970,khoury2004}. Objects are screened
    when their gravitational potential $|\phi_\mathrm{grav}|$ is larger than $\bar{\psi}/\alpha$, where $\bar{\psi}$
    is the cosmic mean of $\psi$ and $\alpha\sim \mathcal{O}(1)$ is the coupling between matter and $\psi$. In
    other words, the effective scalar charge $Q$ that responds to $\psi$ is reduced by the ambient gravitational
    potential. This type of screening operates in $\fr$~\citep{carroll2004, capozziello2003, nojiri2003},
    symmetron~\citep{hinterbichler2010, olive2008, pietroni2005}), and dilaton~\citep{brax2011} theories. In this
    paper we will focus on the Chameleon mechanism within $\fr$, where the Ricci scalar, $R$, in the Einstein--Hilbert
    action is replaced by $R+f(R)$ where $\fr$ is an arbitrary function of $R$.\footnote{
        In practise, the functional form of $\fr$ is tightly constrained by observations~\citep[][]{amendola2007, amendola2008}.}
\item The Vainshtein mechanism, in which the self--interactions of the
    scalar field are determined by the derivatives of $\psi$, which suppress the scalar field and fifth force in
    high density regions~\citep{vainshtein1972, babichev2013}. Scalar fields that exhibit Vainshtein screening
    are generally called `Galileons' because of an internal Galilean symmetry~\citep{nicolis2009}. For an isolated
    spherical source, the force transition happens at a characteristic radius $r_*=(r_s r_c^2)^{1/3}$~(called the
    Vainshtein radius), where $r_s$ is the Schwarzschild radius of the source and $r_c$ in models of interest is
    on the order of the Hubble radius $cH_0^{-1}$.  Within $r_*$ the scalar field is suppressed~($\psi\propto
    \sqrt{r}$), forming a ``sphere'' of screened region around the source. This mechanism is at play in the
    Dvali--Gabadadze--Porrati~\citep[DGP,][]{dvali2000} and massive gravity~\citep{de_rham2011} theories. For
    our purpose, we simplify this class of model as a theory with a $\lcdm$ background cosmology and an extra
    Galileon--type scalar field that manifests Vainshtein screening~\citep{wyman2013}.
\end{itemize}
In both the $\fr$ and Galileon models, the maximum force enhancement is $4/3$ times the normal gravity, but the
``fifth force'' that produces this enhancement has different ranges in the two models. Since the Chameleon scalar field becomes Yukawa--screened,
the fifth force does not have infinite range, i.e., it cannot reach to cosmological scales. Galileons, however,
are never massive, so their force has an infinite range, thus having a much larger impact on linear perturbation
theory than Chameleons do. In the local Universe, however, the Chameleon screening predicts a richer set of
observational signatures that are detectable with astrophysical tests, because it is possible to have order unity
violation of the macroscopic weak equivalence principle~(WEP), i.e., extended objects do not fall at the same rate
as in GR~\citep{hui2009}. In environments of low background $|\phi_\mathrm{grav}|$, objects with deep gravitational
potential can self--screen, while those with shallow potential remain unscreened. For example, \cite{jain2011}
estimated that there could be up to $\sim1\,\kpc$ separation of the stellar
disk~(composed of self--screened objects) from the dark
matter and gas~(both unscreened) inside unscreened dwarf galaxies, using orbital simulations under $\fr$. In contrast,
there is no analogous order one violation in the Vainshtein case, but the Vainshtein ``spheres''\footnote{Strictly
speaking, the Vainshtein mechanism works best for spherically symmetric sources, and not at all for planar sources;
in reality, the screened region always has a complicated geometry that emerges from the geometry of the source.}
of individual objects interfere with each other. For example, in a two--body system where the separation is
$\sim r_*$, the interference reduces the infall acceleration, and this reduction becomes most significant for two
objects with equal masses~\citep{belikov2013}.

The infall zone around clusters lies at the transition between the linear scale, where gravity is universally
enhanced, and the local Universe where GR is frequently recovered, providing a unique avenue for distinguishing MG
from GR. However, in both screening mechanisms the scalar $\psi$ is coupled to density fluctuations via a nonlinear
field equation, which can only be solved jointly with the matter field using numerical simulations. \cite{lam2013}
proposed a halo model-based approach to model the line-of-sight~(LOS) velocity dispersions of galaxies in the infall
region under both modified and normal gravities.  In this study we hope to provide a complete picture of the coherent
motions and distributions of galaxies around clusters in the two MG theories and their systematic deviations from
GR, fully taking into account the nonlinearities that are intrinsic to the Chameleon and Vainshtein mechanisms.

\begin{figure} 
    \centering \resizebox{0.48\figtextwidth}{!} {\includegraphics{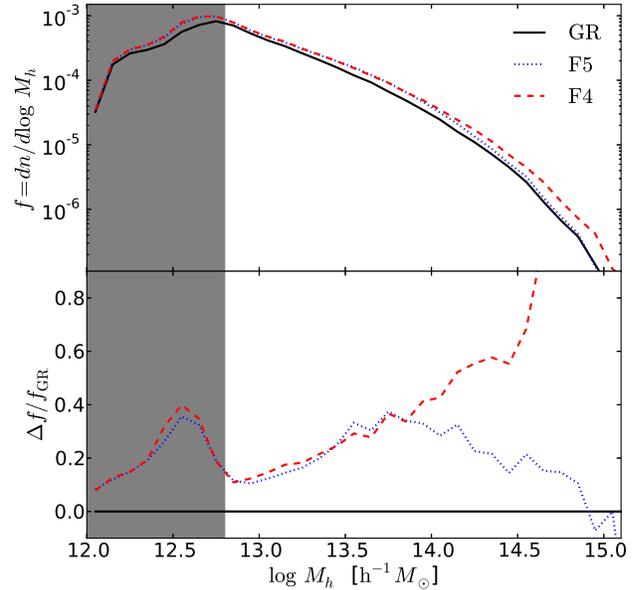}} \caption{{\it Top panel}: Halo mass
    functions in GR~(solid), F5~(dotted), and F4~(dashed) simulations. {\it Bottom panel}: Fractional differences
    in halo mass function between the $\fr$ simulations and the GR simulation. The shaded region in each panel
    indicates the mass scale where the halo catalogs are incomplete.} \label{fig:mf}
\end{figure}

The GIK model of ZW13 describes the average galaxy infall in cluster--centric coordinates in terms of a 2D velocity
distribution at each radius, comprising a virialized component with an isotropic Gaussian velocity distribution
and an infall component described by a skewed 2D {\it t}-distribution with a characteristic infall velocity $\vrc$
and separate radial and tangential dispersions. The virialized component is confined within a ``shock radius''
that is close to the virial radius of the cluster, so in the infall region~(several to $20\,\hmpc$)\footnote{Here
$h\equiv H_0/100\,\mathrm{kms}^{-1}\mathrm{Mpc}^{-1}$} the GIK model reduces to a single infall component. ZW13
demonstrated that GIK profiles can be robustly recovered from the measurement of $\xicgs$ within the Millennium
simulation~\citep{springel2005}, and they applied the method to rich galaxy groups in the Sloan Digital Sky
Survey~\citep[SDSS,][]{york2000}. In particular, the inferred $\vrc$ profile provides a promising way of estimating
the average dynamical mass of clusters and is likely insensitive to baryonic physics and galaxy bias~(Zu et al. in
prep). Although the ZW13 method is calibrated against GR simulations, we will show that the GIK model is also an
excellent description of the infall behavior in MG simulations. Studies have shown that the peculiar velocities are
more distinctively affected by modifications to gravity than the matter density field alone~\citep{wyman2010, jennings2012,
wyman2013, li2013}, and we expect the GIK to be a particularly acute test of MG theories and their screening
mechanisms.

A virtue of using galaxy dynamics in the outskirts of clusters is that independent information on the average
cluster mass profiles can be robustly extracted from stacked weak lensing~(WL) experiments~\citep{mandelbaum2006,
mandelbaum2009, sheldon2009}. Since photons do {\it not} respond to the extra scalar field,\footnote{Though
see \citeauthor{wyman2011} 2011 for an interesting effect on lensing induced by MG} lensing mass estimates will
be different from dynamical mass estimates if gravity is modified from GR on relevant scales~\citep{zhang2007,
reyes2010, zhao2011}. However, implementing this test requires measuring mass profiles on scales where screening
is inefficient, i.e., the cluster infall region rather than within the virial radius. For any given cluster sample
detected by imaging, X-ray, or Sunyaev-–Zel'dovich~(SZ) experiments, one can either directly compare the average
lensing mass and the GIK--estimated dynamical mass or search for inconsistency between the measured $\xicgs$
and the GIK--predicted $\xicgs$ using lensing mass estimates and assuming $\lcdm$+GR. Alternatively, a simplified
test can be performed when WL measurements are unavailable. Since any volume--limited cluster sample is thresholded
by some mass observable~(i.e., galaxy richness, X-ray luminosity, or SZ decrement) that correlates with the true
mass~(with some scatter), we can estimate the corresponding threshold in true mass using the abundance matching~(AM)
technique. However, the uncertainties of AM may be large when the scatter in the mass observable--true mass relation
is large or/and the completeness of the cluster sample is low. For the sake of
simplicity, in this paper we concentrate on 
the AM-based approach, and focus on cluster samples selected to have the same rank order in mass in the GR
and MG simulations.

We present the results of GIK modelling of the Chameleon and Galileon simulations separately in the paper, as the two
simulation sets were run with different initial conditions, cosmic expansion histories, and box sizes and resolution. We first introduce the Chameleon simulations in \S\ref{sec:frsim}, including the specific $\fr$ model, and halo
statistics and kinematics. \S\ref{sec:gik} presents the results of GIK modelling and the measurements of $\xicgs$
for the Chameleon clusters and compares that to GR. In \S\ref{sec:vain}, we presents a similar set of results for
the Galileon clusters. We summarize and discuss the future prospects of our method in \S\ref{sec:con}.

\section{Chameleon Simulations}
\label{sec:frsim}

\begin{figure}
\centering \resizebox{0.48\figtextwidth}{!} {\includegraphics{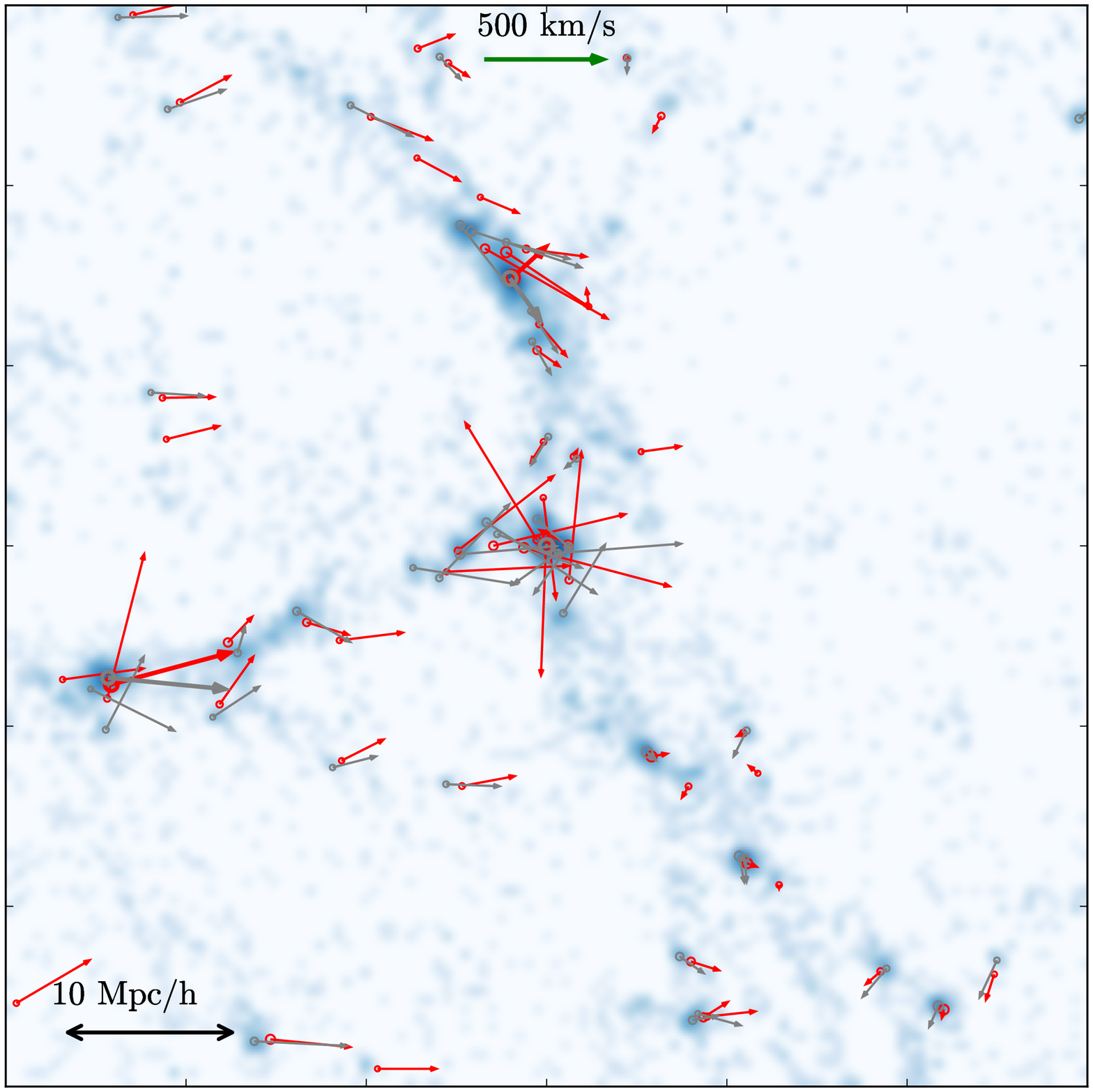}} \caption{Positions and velocities
    ~(in cluster--centric frame) of halos around a massive cluster in the GR and F4 simulations. The dimension of
    the slice is $60\,\hmpc\times60\,\hmpc$ with a
thickness of $30\,\hmpc$, centered on a cluster of $M_h\sim3.2\times10^{14}\,\hmsol$. The length of each arrow is
$v_\mathrm{pec}/H_0$, in units of $\hmpc$. The background grayscale shows the density field of the GR simulation. Gray
and red arrows show the velocities of halos in the GR and F4 simulations, respectively, with larger circles and
thicker arrows for halos of $M>10^{14}\,\hmsol$.} \label{fig:vflow}
\end{figure}

For the Chameleon modified gravity, we use a suite of large--volume $\fr$ simulations~($1\,h^{-3}\mathrm{Gpc}^3$ box
and $1024^3$ particles) evolved with the N-body code ECOSMOG~\citep{li2012}. The same set of
simulations has been used to study the non-linear matter and velocity power spectra~\citep{li2013}, redshift--space
distortions~\citep{jennings2012}, and halo and void properties~\citep{li2012-1} in $\fr$ gravity. The large volume of
the simulations allows us to derive robust statistics from a large number of massive clusters~($M>10^{14}\hmsol$). We
do not make distinctions between the main halos and sub--halos, but include all the bound groups of particles
identified by the spherical density halo finder AHF~\citep[Amiga's Halo Finder,][]{knollmann2009}. The halo
mass is defined by $M\equiv M_\mathrm{vir}=\Delta_\mathrm{vir}\rho_m V_\mathrm{sphere}(r_\mathrm{vir})$, where
$\Delta_\mathrm{vir}$ is the overdensity for virialized halos~\citep[$\Delta_\mathrm{vir}\simeq276$ at $z=0.25$
for typical $\lcdm$ cosmology;][]{bryan1998} and $\rho_m$ is the mean density of the universe.  We will briefly
describe the $\fr$ models here and refer the readers to~\cite{li2012} for more details on the simulations.

\begin{figure*}
\centering \resizebox{0.95\figtextwidth}{!} {\includegraphics{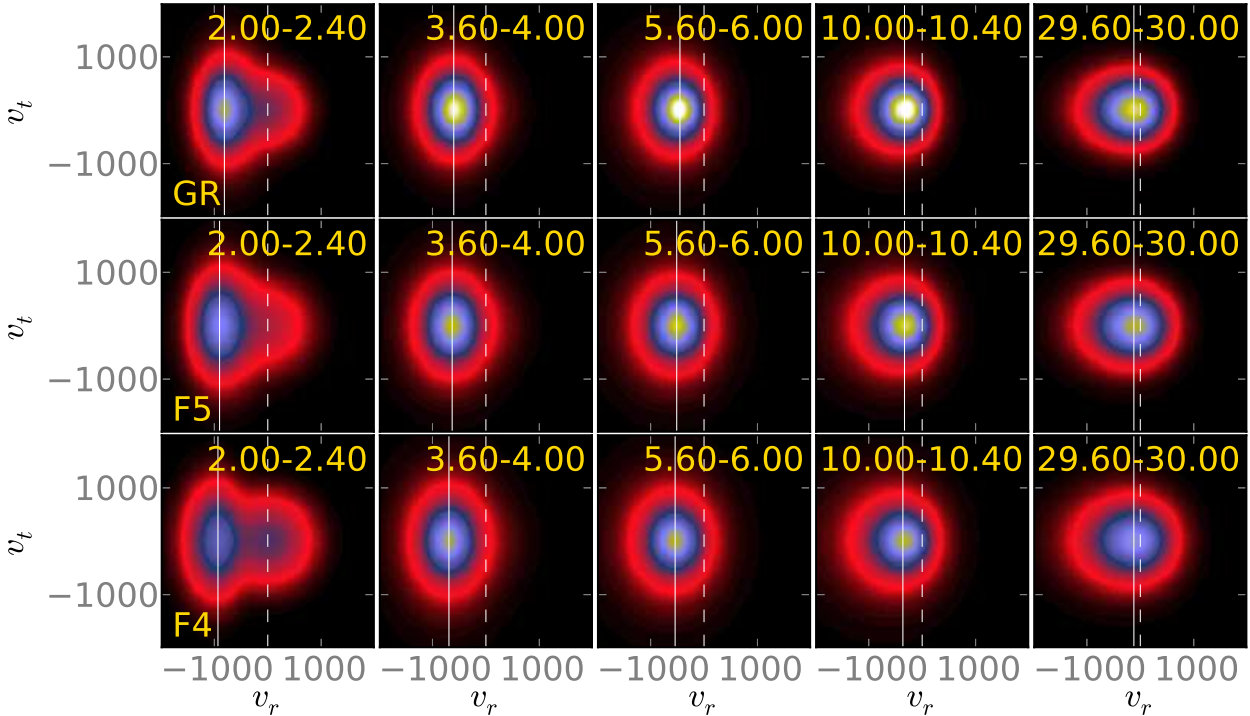}} \caption{Joint probability distributions
of radial and tangential velocities from the best fit to the GR~(top), F5~(middle), F4~(bottom) simulations using
our GIK model, in five different radial bins marked at the top of each panel~(in units of $\hmpc$), for clusters
of $1\text{--}2\times10^{14}\,\hmsol$ in each simulation. In each panel, the vertical dashed line indicates the
position of zero radial velocity while the solid line is the most probable radial velocity. The colour scales,
indicating probability density in the $(v_t, v_r)$ space, are identical across all panels~(colourbars are thus not
shown here).} \label{fig:vblob}
\end{figure*}

The simulations adopt a specific $\fr$ gravity model introduced by~\cite{hu2007}, with the functional form of $\fr$ as
\begin{equation}
    f(R) = -m^2 \frac{c_1(R/m^2)^n}{c_2(R/m^2)^n + 1}, \label{eqn:hsmodel}
\end{equation}
where the mass scale $m^2\equiv H_0^2\om$, $c_1$ and $c_2$ are dimensionless parameters, and $n$ controls
the sharpness of the transition of $\fr$ from $0$ in high--curvature limit~($R$$\to$$0$) to $-m^2 c_1/c_2$ in
low--curvature limit~($R$$\to$$\infty$). The corresponding scalar field in $\fr$ theory is $\frac{d\fr}{dR}$,
commonly denoted as $f_R$~(i.e., the scalaron). Matching to the expansion history of flat $\lcdm$ universes~(e.g.,
$\om=0.24$ and $\ol=0.76$) requires a $R_0/m^2 \sim 41$ and a field value $\f0\simeq-nc_1/c_2^2/(41)^{n+1}<0$, where
the subscript $_0$ represents the present day values. Therefore, the Hu \& Sawicki model is effectively described by
two parameters: $n$ and $\f0$. Models with $|\f0|\lesssim0.1$ are capable of evading solar system tests. For cosmological
tests of Chameleon theories, models with $|\f0|>10^{-4}$ are ruled out by cluster abundance constraints from
\cite{schmidt2009-1}, and models with $|\f0|$ below $10^{-6}$ are nearly indistinguishable from $\lcdm$ universes.
We shall study two cosmologically interesting $\fr$ models with $n=1$ and $\f0=-10^{-5}$, $-10^{-4}$, which will
hereafter be referred to as F5 and F4, respectively. \cite{jain2012} reported constraints of $|\f0|<10^{-7}$ from
other astrophysical tests, so these models may no longer be observationally viable, but they nonetheless provide
useful illustrations of MG effects and a natural comparison for the Vainshtein screening model discussed later.
The expansion history of the two $\fr$ simulations is matched to one flat $\lcdm$ simulation that evolves from the same
initial conditions under normal gravity with $\om=0.24$, $\ol=0.76$, $h=0.73$, and $\s8=0.77$~(referred to as the
``GR'' simulation). The evolution of structure is the same in the three universes up to epochs around $z=49$, which
is the starting time of the simulations, since the fifth force in $\fr$ gravity is vastly suppressed until then.
By studying the time evolution of the matter and velocity divergence power spectra in $\fr$, \cite{li2013} showed
that different $\fr$ models are in different stages of the same evolutionary path at any given time, and that
varying the model parameter $\f0$ mainly varies the epoch marking the onset of the fifth force. The exact epoch
of onset in each model depends on the scale of interest, i.e., at higher redshift for smaller scales~\citep[see
figure 8 in][]{li2013}. We choose to focus on the $z=0.25$ output of the simulations, mimicking the portion of
the Universe most observed by existing and near--term future redshift surveys that contain large samples of clusters.

Figure~\ref{fig:mf} compares the halo mass functions from the GR, F5, and F4 simulations.  The bottom panel shows
the fractional enhancement of the halo mass function in the $\fr$ simulations relative to the GR one.  The shaded
region~($M<M_\mathrm{lim}\equiv6.4\times10^{12}\,\hmsol$) indicates the mass range where the halo catalogs are incomplete
and the bumps in the shaded area of the bottom panel are likely due to numerical effects. Although sub--halos were
included in the halo catalogs, at $M>M_\mathrm{lim}$ the halo mass functions are mostly contributed by main halos. The
two $\fr$ models predict very similar halo abundances below $6\times10^{13}\,\hmsol$, enhanced by $10\text{--}30\%$ over
the GR abundances. For halos in this mass range, the fifth force was activated early enough that the structure
formation has somehow converged in the two $\fr$ models. However, on the high--mass end, the halo mass function in
the F4 model shows even stronger enhancement over GR, while in the F5 model the number of halos becomes closer to
the $\lcdm$ prediction with increasing mass. The divergence of the halo mass functions predicted by the two $\fr$
models beyond $\sim 10^{14}\hmsol$ indicates that the fifth force only started affecting the formation of massive
clusters recently in the F5 model, producing smaller enhancement in the halo abundance compared to F4. For some
very massive clusters in the F5 model~\citep[e.g., see figure 3 in][]{li2012-1}, the gravitational potential
has begun to dominate the background scalaron, activating the Chameleon mechanism to recover GR in the infall
region, while in the F4 model the Chameleon screening likely never activates
anywhere in the universe.  Figure~\ref{fig:mf}
is in good agreement with the study of~\cite{schmidt2009}, where a series of smaller but higher resolution $\fr$
simulations are employed.  \cite{lombriser2013} modeled the halo mass functions measured from the same sets of $\fr$
simulations using environment-- and mass--dependent spherical collapse model in combination with excursion set theory.

\begin{figure*}
\centering \resizebox{0.95\figtextwidth}{!} {\includegraphics{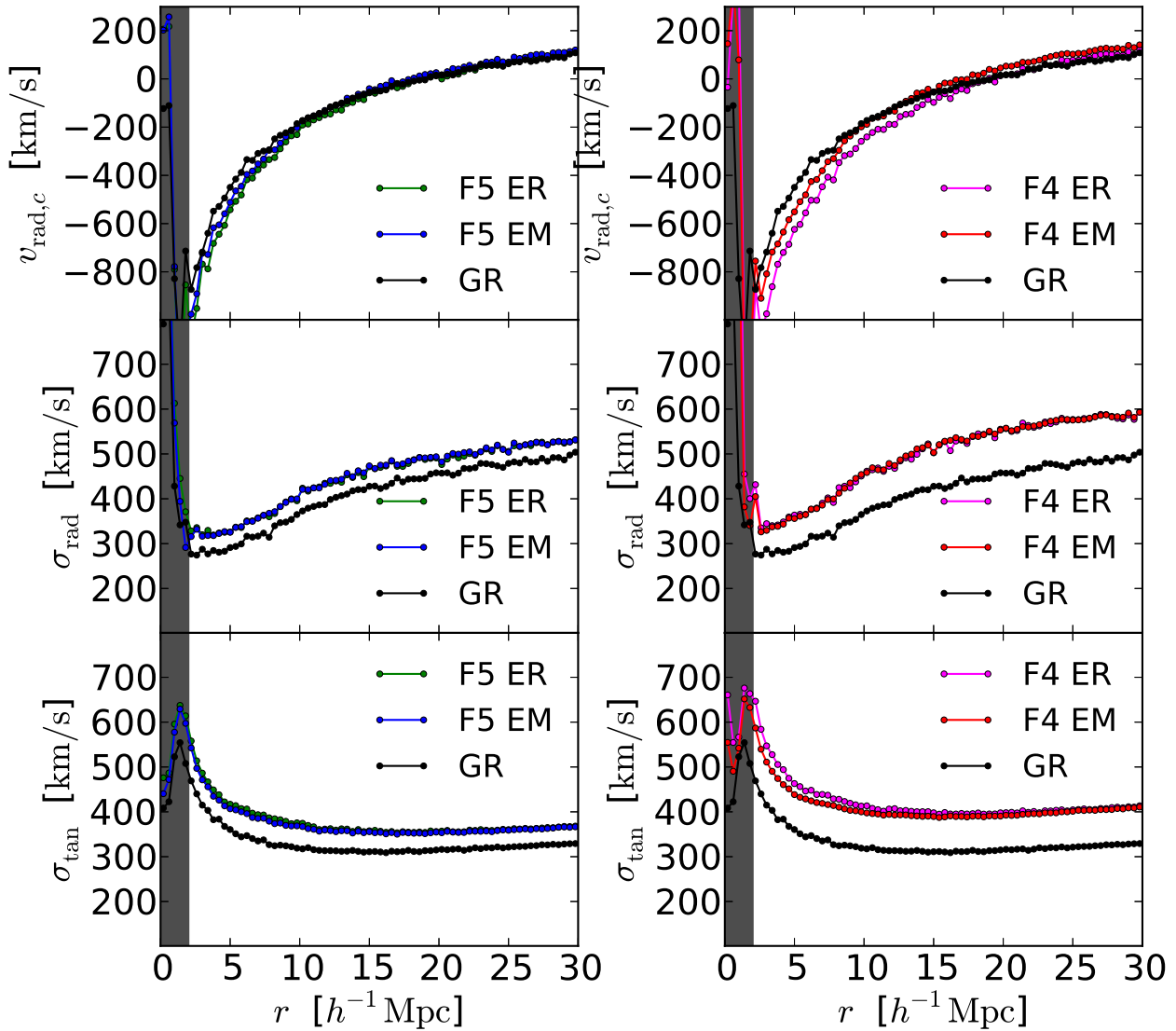}} \caption{Best--fitting characteristic infall
velocity~(top), radial velocity dispersion~(middle), and tangential velocity dispersion~(bottom) as functions
of radius. Left and right columns compare the results of F5 and F4 simulations to that of GR, respectively. The
shaded region in each panel indicates the scales below the maximum infall radius, where the GIK measurements are
less robust. Halos in the $\fr$ simulations are selected as either having equal mass~(EM) or equal rank order~(ER) to
the halos of $1-2\times10^{14}\,\hmsol$ in the GR simulations. See text for details.} \label{fig:v6p}
\end{figure*}

With neither well--resolved sub--halos nor simulated galaxies in the simulations, the common prescription for
constructing mock galaxy catalogs is through Halo Occupation
Distributions~\citep[HODs,][]{jing1998, peacock2000,
seljak2000, scoccimarro2001, berlind2002, zheng2005}. However, the minimum required mass threshold for a complete
halo sample in the Chameleon simulations, $M_\mathrm{lim}$, is overly high for any meaningful HOD --- galaxies
similar to the Milky Way and M31 would be absent. Therefore, we simply use particles and halos as our proxies for
galaxies in the paper, and the behavior of HOD galaxies should be intermediate between that of the ``particle''
galaxies and ``halo'' galaxies. Since we are focused on the relative behavior of MG and GR simulations, the impact
of choices for galaxy proxy on our conclusions should be small. We will usually refer simply to ``galaxies'',
when it is clear from context whether we are using particles or halos as our proxies.

\begin{figure*}
\centering \resizebox{0.95\figtextwidth}{!} {\includegraphics{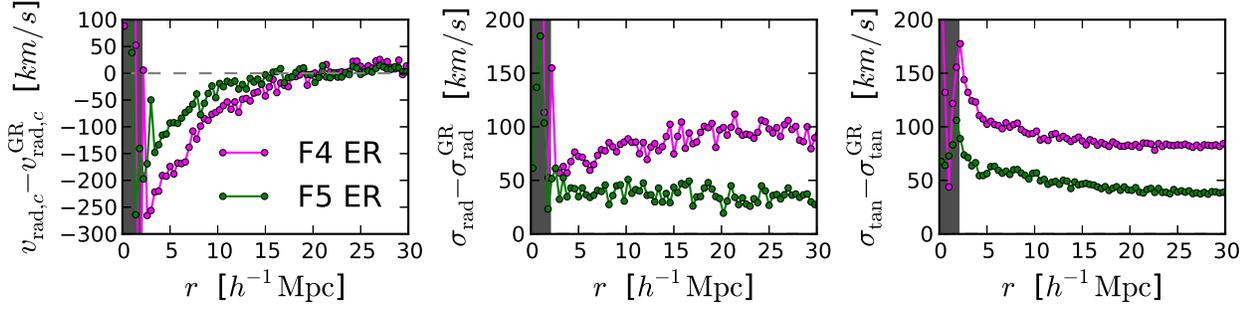}} \caption{Difference in GIK profiles
between the $\fr$ and GR simulations, using clusters of equal rank--order~(ER).} \label{fig:vdiff}
\end{figure*}

Before going into the statistical properties of galaxy infall, we hope to gain
some intuitive understanding of the differences in infall between MG and GR
by looking at Figure~\ref{fig:vflow}. Taking advantage of the same initial
condition shared by the $\fr$ and GR simulations, we locate at the frame
center two primary clusters~($M\sim 3.2\times10^{14}\,\hmsol$) at $z=0.25$
that formed from the same seed in the initial density fluctuation field in the
F4~(red, for which the fifth force is stronger) and GR~(gray) simulations,
respectively, and plot all halos with mass above $6\times10^{12}\hmsol$ as
circles, with their relative velocities to the primary cluster indicated by
the arrows. The dark matter density field in the GR simulation is illustrated
by the grayscale background, highlighting the three filamentary structures
that funnel the infalling halos. The radius of each circle is proportional
to the halo mass, with the thick ones representing halos more massive
than $10^{14}\hmsol$. The length of each arrow is $v_\mathrm{pec}/H_0$,
corresponding to the redshift--space displacement~(in units of $\hmpc$)
that would be seen by a distant observer aligned with the velocity vector.
Figure~\ref{fig:vflow} shows that the
 GIK around {\it individual} clusters is highly anisotropic, and while the
 difference in the spatial distribution
between GR and F4 halos is irregular and fairly mild, it is somewhat enhanced
in redshift space. To avoid clutter, we do not show halos from the F5 model,
which should display smaller differences from GR because of its smaller impact
from the fifth force.  In the next section, we will show that by stacking
individual frames like Figure~\ref{fig:vflow} for clusters of similar mass,
the anisotropy goes away and the average infall kinematics can be well
described by the GIK model proposed in ZW13, for both the $\fr$ and $\lcdm$
models. More importantly, the enhanced difference in the redshift space between
MG and GR models can be captured by systematic differences in the parameters
of the GIK model, namely, the characteristic infall velocity $\vrc$, the
radial velocity dispersion $\sr$, and the tangential velocity dispersion $\st$.

\section{Galaxy Infall Kinematics}
\label{sec:gik}

The halo mass functions in Figure~\ref{fig:mf} suggest that the fifth force became unscreened earlier in F4 than in F5,
which gives it more time to affect the velocity field in the former. As a result, at $z=0.25$ clusters in both
the F4 and F5 models should exhibit more enhanced galaxy infall compared to GR, but we expect the enhancement to
be more substantial in the F4 model. For the small number of screened clusters in the F5 model, the infalling
galaxies around Chameleon clusters should feel similar instantaneous accelerations as their counterparts around
similar clusters in GR. However, the peculiar velocities of galaxies were enhanced by the fifth force when they
were further away from the clusters. By the time they reached to the screened region, the peculiar velocities were
already higher, so the infall stays stronger even though the underlying gravity recovers to GR.  Our goal here is
to quantify this modification to galaxy infall induced by $\fr$ gravities as function of $\f0$ within the framework
of GIK modelling.

To compare the average GIK among the three simulations, we select dark matter halos with
$M_\mathrm{vir}\in1\text{--}2\times10^{14}\,\hmsol$ in the $\lcdm$ simulation as our fiducial GR cluster sample,
and those with the same mass range in the $\fr$ simulations as the ``equal--mass''~(EM) cluster samples. As
mentioned in the introduction, we also select specific halo samples in the $\fr$ simulations to have the
same rank--order in mass as the fiducial GR clusters, i.e., the ``equal--rank''~(ER) cluster samples. The ER
clusters generally have slightly larger masses than the EM ones, with $1.13\text{--}2.14\times10^{14}\,\hmsol$
and $1.27\text{--}2.52\times10^{14}\,\hmsol$ in F5 and F4 models, respectively. The ER sample thus resembles the
set of clusters that formed from the same initial density peaks as the fiducial GR ones. In the limit of very rare,
highly biased peaks, the large--scale cluster bias $b_c$, defined by the ratio between the cluster--matter correlation
function and the matter auto--correlation $\xi_{cm}/\xi_{mm}$, is $\propto\s8^{-1}$, yielding $\xi_{cm}\propto\s8$
on large scales. Operationally, the EM comparison would be most relevant to an observational study of clusters whose
virial masses are calibrated by WL (and thus accurate in both GR and MG). Alternatively, if one ranks clusters by a
mass proxy such as galaxy richness, X--ray luminosity, or SZ signal, then selects clusters above a threshold~(i.e.,
AM method), the ER comparison is more relevant. Hereafter we simply denote the fiducial GR sample as ``GR'' while
comparing it to the ``EM'' and ``ER'' samples in MG simulations.

\subsection{Velocity Field Model}
\label{sec:vel}

\begin{figure*}
\centering \resizebox{0.95\figtextwidth}{!} {\includegraphics{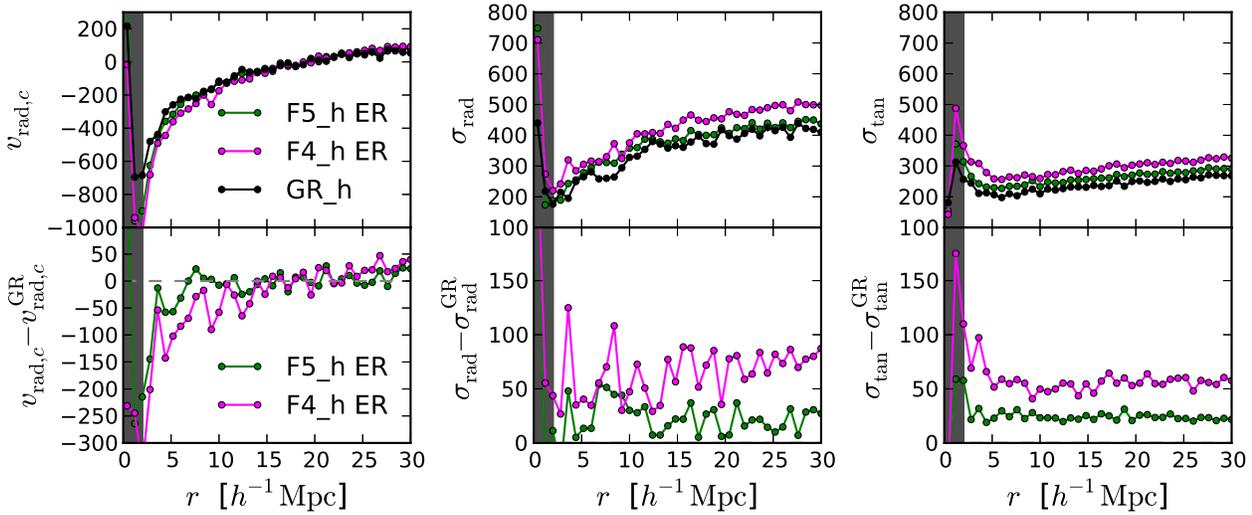}} \caption{Comparison of GIK profiles
in the GR, F5, and F4 simulations, using halos as proxy for galaxies instead of dark matter particles as in
Figures~\ref{fig:v6p} and~\ref{fig:vdiff}.} \label{fig:vdiff_h}
\end{figure*}

As mentioned in the introduction, the two--component GIK model is an excellent analytic description of the joint
2D distribution of radial and tangential velocities of galaxies in the cluster--centric frame, $\p2d$. Here we
will present the results of GIK modelling for the cluster samples defined above, and refer the readers to ZW13 for
details on the GIK parameterization and fitting procedures.

Figure~\ref{fig:vblob} shows the best--fitting $\p2d$ for the GR~(top), F5--EM~(middle), and F4--EM~(bottom)
cluster samples at five different radial bins, using dark matter particles as proxy for galaxies. Negative $v_r$
indicates falling toward clusters. Following ZW13, we define $v_t$ as the tangential velocity component that
is projected in the plane of LOS axis and galaxy position vector in the cluster--centric frame~(see the 3D
diagram in the figure 2 of ZW13). Since the {\it average} galaxy motion around the cluster center is isotropic,
the probability distribution of $v_t$ is symmetric about zero. Hereafter we refer to $v_t$
simply as the ``tangential velocity''. Note that the Hubble flow is subtracted when defining $v_r$, but it will
be incorporated when modelling $\xicgs$. The GIK of the three cluster samples in Figure~\ref{fig:vblob} show some
generic trends with radius: 1) The distribution has two distinct components on very small scales~(leftmost column)
but only shows a single infall component on large scales; 2) the infall component is symmetric about the mean $v_r$
near the turn--around radius, where infall velocity is comparable to Hubble flow~(middle column); 3) the $v_r$
distribution of the infall component is skewed toward positive velocities beyond the turn--around radius~(two
right columns), but is negatively skewed below that radius~(two left columns).

The impacts of modified gravity on $\p2d$ are subtle but nonetheless visually apparent in Figure~\ref{fig:vblob}
when comparing the three models at the same radial bin~(i.e., within the same column). The solid vertical line
in each panel indicates the most probable radial velocity, which shifts to more negative $v_r$ with increasing
$|\f0|$~(i.e., from top to bottom) at each radius. Simultaneously, the dispersions of the infall component in the radial
and tangential directions also increase as function of $\f0$ --- the joint distributions in the F4 model are more
extended than in GR. This increased width results in the decreased peak amplitude of the distributions (which are
normalized to unity by definition).

To quantify the differences in GIK among the three simulations, we will focus on the impacts of modified gravity on
three of the GIK parameters~($\vrc$, $\sr$, and $\st$).  We ignore the other GIK parameters defined by ZW13~(seven
in total), including two parameters describing the virialized component that are irrelevant to this paper, and
two others that describe the skewness and the kurtosis of the infall component, which we found to be insensitive
to modified gravity. Note that the characteristic radial velocity $\vrc$ is not the mean, median, or mode of the
radial velocity distribution, but is a characteristic velocity naturally associated with the definition of the
skewed {\it t}-distribution used for describing the GIK infall component~(see ZW13, equation 6).

Figure~\ref{fig:v6p} presents the best--fit GIK parameters as functions of radius for the F5~(left) and the
F4~(right) models, respectively. In each panel we show the parameter profiles for all three types of cluster
samples~(GR, EM, and ER). The shaded region in each panel indicates the radial bins where GIK is dominated by
the virialized component and the fit to the infall component is less robust.  For the F5 model, the halo mass
function is close to that of $\lcdm$ at high masses, so the EM and ER samples have little difference in mass and
their GIK profiles look very similar. However, as expected from Figure~\ref{fig:vblob}, the two F5 cluster samples
in the $\fr$ simulation show stronger infall~(top left) and larger velocity dispersions~(center and bottom left)
compared to the GR sample. For the F4 model, the difference between the EM and ER samples is larger, especially in
the $\vrc$ profiles~(top right), where the difference between the ER and GR profiles is almost double that between
the EM and GR ones. For both samples, the F4 results are more easily distinguished from GR than the F5 results,
as expected. Since we anticipate that ER comparisons will be more observationally relevant in most cases, we will
focus henceforth on the GIK of ER cluster samples in the MG simulations.

We highlight the differences in GIK profiles between ER and GR clusters for both Chameleon models in
Figure~\ref{fig:vdiff}. The characteristic infall velocity profiles exhibit significant effects from the Chameleon
gravity on scales below $15\,\hmpc$, showing $\sim 100\,\kms$ and $200\,\kms$ enhancement at $5\hmpc$ for the F5
and F4 models, respectively. Beyond $15\,\hmpc$ the $\vrc$ profiles converge to the GR prediction.  The dispersion
profiles in $\fr$ models deviate from GR on all distance scales, with the differences almost constant and decreasing
with radius for $\sr$ and $\st$, respectively. The magnitude of the deviations we see here is very encouraging ---
$\sim 100\,\kms$ difference in both the $\vrc$ and the dispersions is already detectable within $2\sigma$ in ZW13,
where a preliminary GIK constraint is obtained using two samples of SDSS rich groups~(with group number $\sim 2000$
and $\sim 600$, respectively) and the SDSS DR7 main galaxy sample.

Using dark matter particles as proxy for galaxies effectively assumes that galaxies have the same density profile
and velocity distribution as dark matter particles within halos. In reality, we expect central galaxies to have
low peculiar velocities relative to the halo center of mass, and the spatial distribution may be less concentrated
than the matter~\citep[see, e.g.,][]{budzynski2012}. To bracket the expectations for constraining GIK using realistic
galaxy samples, we repeat the above experiment using halos instead of particles as proxy for galaxies. This mimics the
scenario where a Luminous Red Galaxy~\citep[LRG,][]{eisenstein2001} galaxy sample is employed, implying approximately
one galaxy per halo as one extreme of the HOD~\citep{zheng2009}. Figure~\ref{fig:vdiff_h} summarizes the result
of this experiment. We denote the curves correspondingly as ``\texttt{GR\char`_h}'', ``\texttt{F5\char`_h}'',
and ``\texttt{F4\char`_h}'' in the figure. The GIK profiles are much noisier because of the rarity of halos,
but the differences among the three samples are similar to those seen in Figure~\ref{fig:vdiff}, but smaller
in magnitude by about a factor of two. We infer that the difference in GIK seen in Figure~\ref{fig:vdiff} using
``particle'' galaxies has approximately equal contributions from two sources, Chameleon modifications to the random
motions within halos~(i.e., ```1-halo'') and the impact of Chameleon gravity on the bulk inflow of halos~(i.e.,
```2-halo''). Quantitative predictions for a particular galaxy sample will require simulations that resolve the
host halos and thus allow a full HOD model of the population, incorporating both 1-halo and 2-halo effects with
appropriate weight.

\subsection{Cluster--galaxy Correlation Function}
\label{sec:corr}

\begin{figure*}
\centering \resizebox{0.95\figtextwidth}{!} {\includegraphics{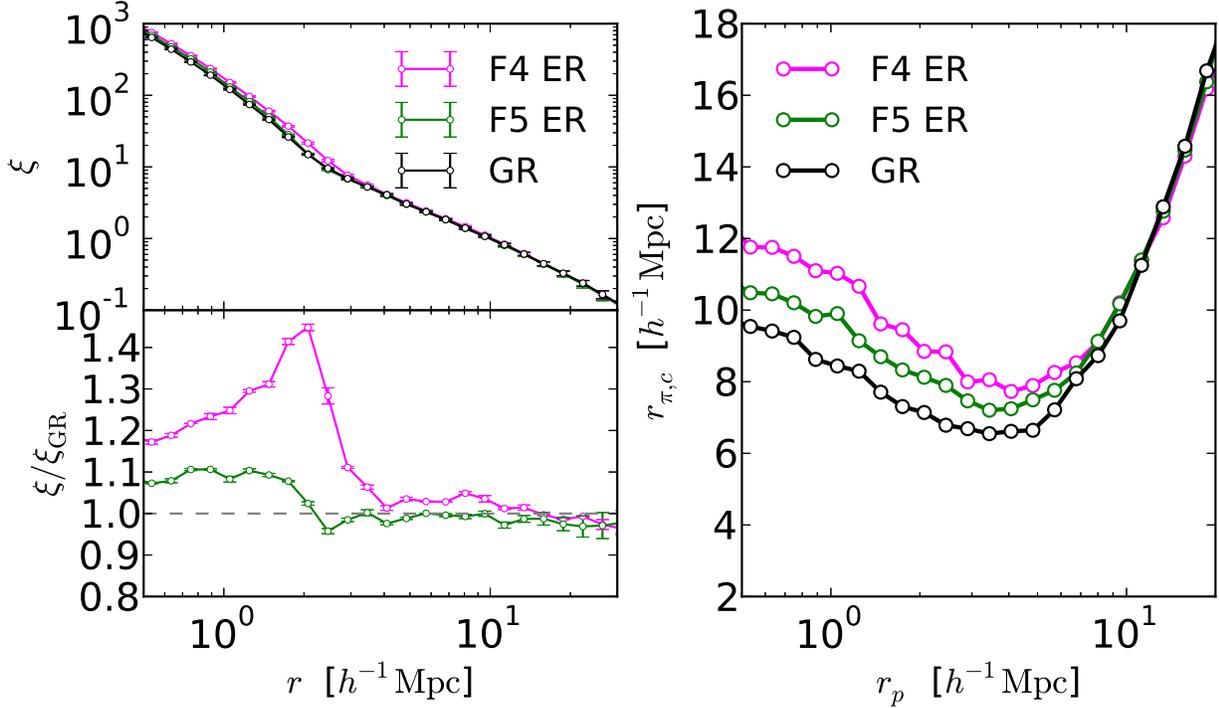}} \caption{{\it Left Panel}: Comparison
    of real--space correlation functions in the GR, F5, and F4 simulations, using particles as proxy for galaxies. {\it
    Right Panel}: Characteristic LOS distances
derived from the redshift--space correlation functions in Fig.~\ref{fig:xirs}.} \label{fig:xi3d_ucurve}
\end{figure*}

The redshift--space cluster--galaxy cross--correlation function, $\xicgs$, is a comprehensive characterization
of the statistical relation between clusters and galaxies, influenced by both the real--space cross--correlation
$\xicgr$ and the peculiar velocities induced by the cluster gravitational potential. Mathematically, $\xicgs(r_p,
r_\pi)$, a function of projected cluster--galaxy separation $r_p$ and line-of-sight redshift separation $r_\pi$,
can be derived by convolving the real--space $\xicgr(r)$ with the Hubble flow--corrected LOS velocity distribution,
which can be straightforwardly predicted from the GIK model~(see equation 11 in ZW13). ZW13 demonstrated that the GIK
can be extracted from the observed $\xicgs$, by taking advantage of the non--degenerate imprint of each GIK element
on the 2D pattern of $\xicgs$.  In this section we will examine the impact of Chameleon gravity on the real and
redshift--space cluster--galaxy correlation functions in the $\fr$ simulations, using particles as proxy for galaxies.

For measuring $\xicgr$, we count the numbers of particles around clusters in spherical shells of successive
radii, ranging from $100\,\hkpc$ to $30\,\hmpc$ with logarithmic intervals, then average over all clusters in
each bin and normalize by the particle numbers expected in a randomly located shell of equal volume. We measure
$\xicgs$ in a similar way, counting galaxies in cylindrical rings of successive LOS distance $r_\pi$ for each
projected separation $r_p$~(assuming a distant--observer approximation so that the LOS is an axis of the box).
Uncertainties in both measurements are estimated by Jackknife re--sampling the octants of each simulation box.

We start by showing the real--space cluster--galaxy correlation function $\xicgr$ for the GR, F4--ER, and F5--ER
cluster samples in the left panels of Figure~\ref{fig:xi3d_ucurve}. In the top left panel, all three correlation
functions exhibit a break at $2-3\,\hmpc$, marking the transition from the NFW--like density
distribution~\citep{navarro1997} within halos to a
biased version of the matter auto--correlation function on large scales~\citep{hayashi2008,zu2012}. The bottom left
panel shows the ratio of $\xicgr$ between the ER samples in $\fr$ simulations and the GR sample. On scales below the
break radius, $\xicgr$ of the ER clusters show enhancement of $10\%$ and $20\%$ in the F5 and F4 models, respectively,
because they are intrinsically more massive than their counterparts in the $\lcdm$ simulation. On scales larger
than $\sim 5\,\hmpc$, the ER clusters have nearly the same large--scale clustering as the GR sample.  Since we
are measuring the cross--correlation with dark matter particles, these $\xicgr$ profiles also determine the
cluster--galaxy WL profile~(see, e.g., eq. 13 of \citeauthor{zu2012} 2012).

On intermediate scales, there is a bump at $\sim 2\,\hmpc$ in the ratio between $\xicgr$ of the F4--ER and the
GR samples, but not in the ratio curve for F5--ER. In the F4 model, the fifth force became unscreened from quite
early time, so that the velocity field has been enhanced for a long time by $z=0.25$. This means that the peculiar
velocities are enhanced by roughly the same factor $\kappa$ as the gravitational force, where $\kappa$ is between
unity and $4/3$, and the kinetic energy of particles in F4 is thus $\kappa^2$ times that in GR; meanwhile, the
gravitational potential in F4 is $\sim\kappa$ times as deep as in GR.  The net result is that the MG effect on the
particle kinetic energy dominates over its effect on the cluster potential, so that as a compromise the particles
tend to move toward the outer parts of clusters. The situation is different in the F5 model, where the fifth force
became unscreened quite late. By $z=0.25$, the fifth force has become unscreened but only for a short period, so
that particle velocities have not been significantly affected by it. On the other hand, due to the disappearance
of the screening, the potential of the cluster suddenly became deeper. The result is a stronger MG effect on the
potential than on the kinetic energy of particles, and particles tend to move toward the inner parts of clusters. These
features have been observed in the halo density profiles of $\fr$ simulations before~\citep[e.g.,][]{zhao2011-1}, and
similar ones have been found in coupled quintessence simulations. Using cluster WL measurements, \cite{lombriser2012}
exploited this small--scale enhancement of cluster density profiles in Chameleon gravity and
obtained a constraint of $|\f0|<3.5\times10^{-3}$ at $95\%$ confidence level.

\begin{figure*}
\centering \resizebox{0.95\figtextwidth}{!} {\includegraphics{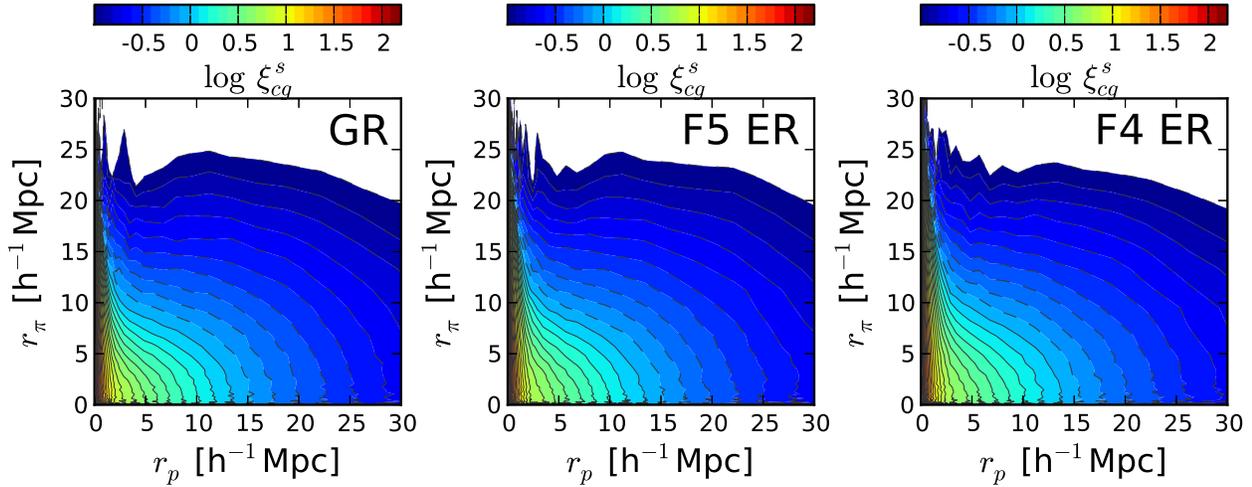}} \caption{Redshift--space correlation functions
in the GR, F5, and F4 simulations.} \label{fig:xirs}
\end{figure*}

Figure~\ref{fig:xirs} presents the redshift--space cluster--galaxy correlation functions for the three cluster
samples, showing a stronger small--scale Fingers-of-God~\citep[FOG,][]{jackson1972} effect with increasing $|\f0|$,
but with similar LOS squashing effect on large scales~\citep[a.k.a., Kaiser effect,][]{kaiser1987}. However, the
most easily visible features of $\xicgs$ in Figure~\ref{fig:xirs} are driven by the radial gradient of cluster
density profiles, which are fairly insensitive to the influence of Chameleon modifications to gravity according to the
left panel of Figure~\ref{fig:xi3d_ucurve}. Modified gravity also changes the shape of $\xicgs$ at
fixed $r_p$ via its effects on the GIK, re--distributing matter/galaxies along the $r_\pi$ axis. To reveal
this LOS distortion of $\xicgs$ by $\fr$, following ZW13, we compute the characteristic LOS distance $\rpic(r_p)$
by fitting a powered exponential function to $\xicgs$ at each fixed $r_p$,
\begin{equation}
\xicgs(r_p, r_\pi) \simeq \xicgs(r_p, r_\pi=0) \exp \left\{-\left|\frac{r_\pi}{\rpic}\right|^\beta\right\},
\label{eqn:pe}
\end{equation}
where $\rpic$ is the characteristic length scale at which $\xicgs$ drops to $1/e$ of its maximum value at
$r_\pi=0$. The shape parameter $\beta$ yields a Gaussian cutoff for $\beta=2$ and simple exponential for
$\beta=1$, though any value is allowed in the fit. The results of this fitting are shown in the right panel of
Figure~\ref{fig:xi3d_ucurve}. The $\rpic$ vs. $r_p$ curves exhibit the characteristic U-shape discovered in ZW13
--- FOG stretching at small $r_p$ gives way to Kaiser compression at intermediate $r_p$ which gives way to Hubble
flow expansion at large $r_p$. Clearly, the $\xicgs$ distribution along the LOS is a sensitive probe of $\fr$
models at $r_p<6\,\hmpc$, becoming more extended with increasing $|\f0|$~(e.g., $\rpic=7$, $8$, and $9\,\hmpc$
at $r_p=2\,\hmpc$ for GR, F5, and F4 models, respectively).

The detailed shape of $\xicgs(r_p, r_\pi)$ reflects a complex interplay among the four elements of the galaxy
kinematics around clusters, including the three GIK profiles of the infall component~($\vrc$, $\sr$, and $\st$) and
the virial component~(see figure 10 of ZW13 for an illustrative experiment). For the same reason, the increase of
$\rpic$ with $\f0$ has different origins at different projected distances. Below $r_p\sim1.5\,\hmpc$, the response
of $\rpic$ to $\f0$ is uniform with $r_p$, caused by the uniform increase of dispersion in virial motions. For
$r_p \sim 1.5-3\,\hmpc$, the $\f0$ dependence of $\rpic$ has two contributing sources, one being the increase of
tangential velocity dispersions, the other the increase of maximum infall velocities, which transport high--speed
matter/galaxies from one side of the cluster in real space to the opposite side in redshift space~(i.e., the portion
of the FOG effect caused by infall). For $r_p \sim 3-6\,\hmpc$, $\f0$ influences the $\rpic$ profile mainly via the
increase of tangential velocity dispersions. The radial velocity dispersions only enter into play at large $r_p$,
where the diagnostic power of $\xicgs$ is diminishing.

% Mark's simulations
\section{Results for a Galileon Model}
\label{sec:vain}

\begin{figure*}
\centering \resizebox{0.95\figtextwidth}{!} {\includegraphics{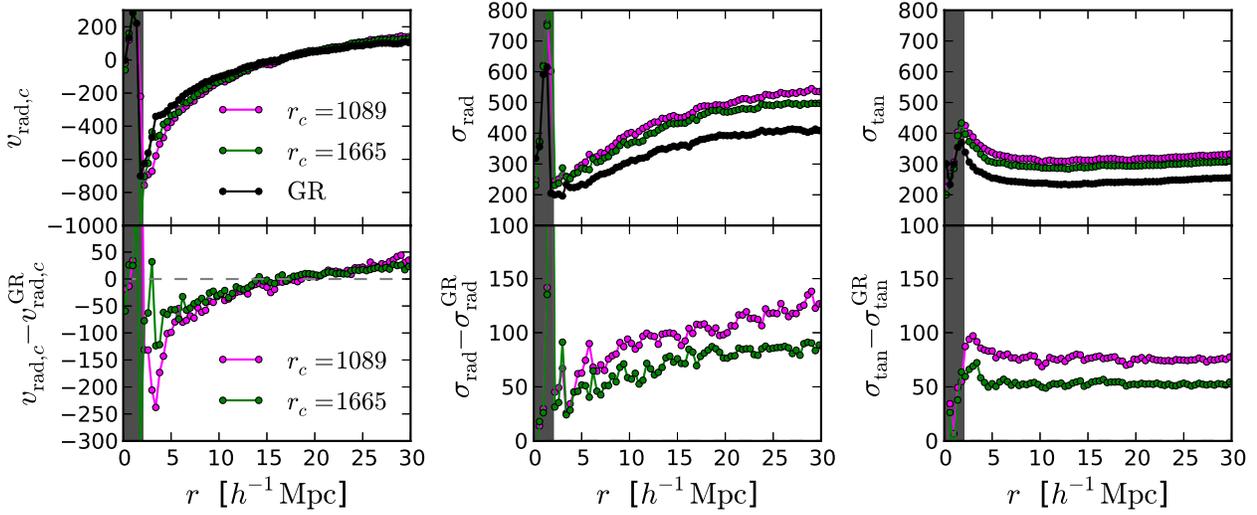}} \caption{Similar to
Figure~\ref{fig:vdiff_h}, but for the Galileon simulations and using particles as proxy for galaxies, around
clusters of equal rank order~(ER).
}
\label{fig:vdiff_mark}
\end{figure*}

Although Chameleon theories like the Hu \& Sawicki $\fr$ model include the phenomenology of $\lcdm$ without a
{\it true} cosmological constant, \cite{wang2012} proved that the theories that invoke a Chameleon--like scalar to
explain cosmic acceleration essentially rely on a form of dark energy rather than a genuine MG effect,\footnote{For
a ``genuine'' MG effect, the cosmic acceleration should stem entirely from the conformal transformation from
the Einstein frame to the Jordan frame. See~\cite{wang2012} for details.} even if they are initially described in
terms of an altered gravitational action. Conversely, the Galileon class of theories is capable of accelerating
the cosmic expansion even in the absence of any form of dark energy~\citep[e.g.,][]{de_rham2011-1, gratia2012,
appleby2012, barreira2012}, i.e., they have so--called ``self--accelerating'' solutions. Here we study the GIK for
a simplified version of such kind of Galileon theories, where an extra Galileon--type scalar field that manifests
the Vainshtein mechanism permeates a universe with the $\lcdm$ background cosmology.

We employ a suite of Galileon simulations with $512^3$ particles on a $512^3$ grid of $L_\mathrm{box}=400\hmpc$,
evolved using the Particle Mesh code of~\cite{khoury2009}, which was updated
by~\cite{wyman2013}~\citep[for other
Galileon/DGP simulations, see, e.g.,][]{chan2009, schmidt2009-2, li2013-1, barreira2013}. The simulations were first
used by~\cite{wyman2013} for studying the statistics of matter clustering in real and redshift spaces. We will briefly
introduce the Galileon implementation and parameterization here and refer the readers to \cite{wyman2013} for details.

As described in the introduction, the Vainshtein mechanism has only one parameter, $r_c$, which is interpreted
as the Compton wavelength associated with the graviton mass in massive gravity theories, so that the Vainshtein
radius of a point mass with mass $M$ is
\begin{equation}
r_*^p\equiv(r_s r_c^2)^{\frac{1}{3}}\simeq2.1 \left(\frac{M}{10^{14}M_\odot}\right)^{\frac{1}{3}}
\left(\frac{r_c}{10^{3}\mpc}\right)^{\frac{2}{3}}\mpc.  \label{eqn:rstar}
\end{equation}
For extended objects of the same mass $M$, the Vainshtein radii are generally several times larger
than $r_*^p(M)$ --- e.g., a NFW halo with $M=10^{14}M_\odot$ has $r_*\sim10\,\mpc$, and galaxies generally
have $r_*\sim 1\,\mpc$. Therefore, whereas the cluster interior below the virial radius belongs to the strongly
Vainshtein--screened regime, the cluster infall region is weakly screened, displaying complex interference among
Galileon fields sourced by the primary cluster and the infalling galaxies and galaxy groups.

We make use of three simulations with the same expansion history and initial condition, one of them a flat
$\lcdm$ universe evolved under normal gravity with $\om=0.24$, $\ol=0.76$, $h=0.73$, and $\s8=0.80$, and the
other two immersed in Galileon scalar fields with $r_c=1665$ and $1089\,\mpc$. In the figures we refer to them
simply as ``GR''~(i.e., $r_c=+\infty$), $r_c=1665$, and $r_c=1089$, respectively, and a smaller $r_c$ implies
an earlier onset of the fifth force and an effectively stronger fifth force in the late Universe~\citep[see,
e.g., figure 5 of ][for the change of linear growth rate as function of $r_c$]{wyman2013}.  Note that the
``GR'' simulation is different from what we used for comparing to the $\fr$ simulations, albeit with similar
cosmology. Dark matter halos are identified via a spherical overdensity finder with the halo mass defined by
$M\equiv M_{200}=\Delta_{200}\rho_m V_\mathrm{sphere}(r_{200})$, different from the $M_\mathrm{vir}$ used for the
Chameleon simulations. Since $\Delta_\mathrm{200} < \Delta_\mathrm{vir}$, a halo would have a higher $M_{200}$
than $M_\mathrm{vir}$~(e.g., $M_{200}\sim 1.1\times M_\mathrm{vir}$ for a $10^{14}\hmsol$ cluster at $z=0.25$
in $\lcdm$). The halo mass functions, halo bias functions, and matter power spectra of the three simulations
can be found in \cite{wyman2013}. Targeting the same redshift range as in \S\ref{sec:frsim}, we use the $z=0.20$
output of the simulations~(not $z=0.25$ due to the different sets of recorded epochs in the two suites of simulations).

Since the volume of the Galileon simulations is only $6.4\%$ of that of the
Chameleon simulations, we have to select samples with a wider mass bin size for robust GIK measurements. For the
fiducial cluster sample in the GR simulation we include clusters with $M_{200}\in1\text{--}3\times10^{14}\,\hmsol$,
and similar to \S\ref{sec:frsim} we also select two ER cluster samples from respective Galileon simulations, with
$M_{200}\in 1.15\text{--}3.52 \times10^{14}\,\hmsol$ and $M_{200}\in1.23\text{--}3.87\times10^{14}\,\hmsol$ in the
$r_c=1665$ and $r_c=1089$ models, respectively. Unlike \S\ref{sec:frsim}, we do not show the results from the EM
cluster samples for the GR vs. Galileon comparisons, as the relative difference between the EM and ER samples is similar
to what we see in the Chameleon simulations. Because of the smaller volume, we can only afford to use dark matter
particles as proxy for galaxies. Note that the Galileon cluster sample here comprises halos that are intrinsically
smaller than the one used in~\S\ref{sec:frsim}, due to different overdensity thresholds used in mass definitions.

Figure~\ref{fig:vdiff_mark} compares the GIK profiles of the two ER samples in the Galileon simulations to that of
the fiducial GR cluster sample.  As intrinsically less massive systems, the fiducial GR clusters show weaker infall
velocities and velocity dispersions than the GR clusters used in the Chameleon comparison at all distances~(e.g.,
comparing the black curves in the top panels of Figure~\ref{fig:vdiff_mark} to the black curves in Figure~\ref{fig:v6p}). However, the relative
difference between the MG and GR samples overall looks very similar to what we see in the Chameleon comparison~(e.g.,
comparing the bottom panels of Figure~\ref{fig:vdiff_mark} to Figure~\ref{fig:vdiff}).  As expected, $\vrc$, $\sr$,
and $\st$ all become stronger with decreasing $r_c$~(i.e., stronger fifth force). Specifically, $\vrc$ shows $\sim
50$ and $100\,\kms$ enhancement at $r=5\,\hmpc$ for the $r_c=1665$ and $r_c=1089$ models, respectively, which are
comparable to the fractional enhancements in the F5 and F4 Chameleon models, respectively.

The only major difference between Figure~\ref{fig:vdiff_mark} and Figure~\ref{fig:vdiff} appears in the bottom middle
panel: The deviation of $\sr$ profiles from the GR prediction decreases as function of distance, from $125$/$90\,\kms$
at $r=30\,\hmpc$ to $60$/$50\,\kms$ at $r=5\,\hmpc$, whereas in the Chameleon comparison the deviation of $\sr$
stays more or less constant with distance. This difference in $\sr$ may be reflecting the different ranges of
fifth force in the two models: the Galileon has infinite range, so the force is enhanced further way from clusters,
whereas the Chameleon force is Yukawa--suppressed on large scales. However, the pattern of $\xicgs$ is insensitive
to $\sr$ at $r_p<6\,\hmpc$ where the $\xicgs$ measurement is the most robust, making it an unpromising tool for
distinguishing the two mechanisms. The GIK profiles of the EM cluster samples in the Galileon simulations closely
follow those of their corresponding ER counterparts, albeit with slightly weaker amplitudes.

\begin{figure*}
\centering \resizebox{0.95\figtextwidth}{!} {\includegraphics{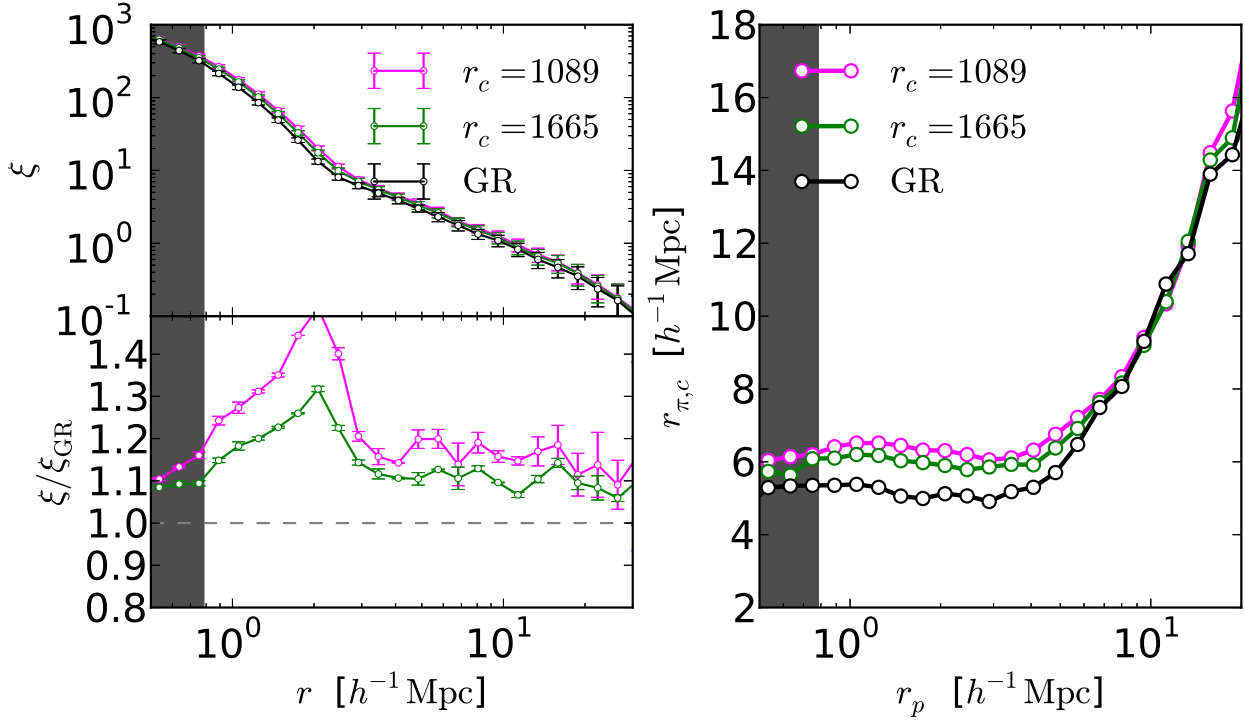}} \caption{Similar to
Figure~\ref{fig:xi3d_ucurve}, but for the Galileon simulations. The shaded region indicates the distance scale
below the force resolution in the simulations.} \label{fig:xi3d_ucurve_mark}
\end{figure*}

Figure~\ref{fig:xi3d_ucurve_mark} compares the real--space cluster--galaxy cross--correlation function $\xicgr$~(left)
and the characteristic LOS distance $\rpic$~(right) measured for the three cluster samples. The shaded region
indicates the scales below the force softening length of the simulations, where the correlation function and
velocity dispersions are artificially suppressed.  On small scales, the ratios between $\xicgr$ of the Galileon ER
samples and fiducial GR sample ~(bottom left) display similar features to these in Figure~\ref{fig:xi3d_ucurve},
including an enhancement interior to the virial radius because the ER clusters are more massive, and a bump around
$2\,\hmpc$, though this is only $\sim 1.2\,\hmpc$ away from the force resolution limit. To check whether the bump
is a numerical artefact, we repeated the same measurements of $\xicgr$ using a suite of higher resolution~(but
smaller volume) Galileon simulations and verified that the bump is physical. Similar to the $\fr$ case in the F4
model, the fast transition from Galileon force outside to normal gravity inside causes a sudden change of the depth
of the potential well, but not the galaxy velocity (or kinetic energy), which has been experiencing enhancement
well before infall. As the galaxies have excessive kinetic energy, they tend to move to the outer parts of halos,
making the density profile lower in the central region of clusters and higher near the edges.

On large scales, the ER samples in the Galileon simulations exhibit stronger clustering than the GR clusters. This
enhancement in $\xicgr$ can be understood by starting from the findings of \cite{wyman2013}, which suggest that
the halo mass function, the halo bias function, and the matter power spectrum of the Galileon simulations are like
those of $\lcdm$ universes with higher $\s8$.\footnote{Though they can be distinguished from $\lcdm$ by examining the
    Kaiser effect, where changing $\s8$ cannot mimic the large scale boost in the redshift--space clustering due
    to MG. See~\cite{wyman2013} for details.}
For example, at $z=0$ the $r_c=1665$ and the $r_c=1089$ simulations resemble the $\lcdm$ universes with $\s8=0.88$
and $\s8=0.92$, respectively. Since the large--scale $\xicgr$ of ER clusters is almost linearly
proportional to the effective $\s8$, we observe $\sim10\%$ and $15\%$ enhancement
in $\xicgr$ at $\sim10\,\hmpc$ for clusters in the $r_c=1665$ and $r_c=1089$ models, respectively.

For the redshift--space cluster--galaxy cross--correlation function, because the impact of the Galileon field on
GIK is similar to that in the Chameleon models, the $\rpic$ curves in Galileon simulations also exhibit similar
deviations from the GR curves, as shown in the right panel of Figure~\ref{fig:xi3d_ucurve_mark}. The lack of upturn
of $\rpic$ at small $r_p$ is a consequence of force resolution suppressing velocity dispersions; we expect that
simulations with higher force resolution would show the characteristic U-shape for all three models.  The overall
lower amplitude of $\rpic$ compared to Figure~\ref{fig:xi3d_ucurve} is again the result of selecting intrinsically
less massive halos. When we select the Galileon clusters to have equal--mass to the GR clusters, the enhancement
in large--scale clustering~(left panel) disappears because the large--scale
bias is steep function of mass, but
the differences in GIK and $\rpic$ remain.

\section{Conclusions}
\label{sec:con}

% some thoughts: 
% \begin{itemize}
    % \item main conclusion, it's possible!
    % \item whether it is generic or coincidence that the impacts are similar?
    % \item the how to distinguish the two?
    % \item how it compares with Lam et al. paper
    % \item future combinations of spectroscopic and imaging surveys
% \end{itemize}

We have investigated the impact of modified gravity on the galaxy infall motion around massive clusters by applying
the GIK model developed in ZW13 to two suites of $\fr$ and Galileon N-body simulations. Both MG theories seek
to explain cosmic acceleration by modifying GR on cosmological scales, but they recover GR in dense regions via
two distinct ``screening'' effects: the potential--driven Chameleon mechanism in $\fr$ and the density--driven
Vainshtein mechanism in Galileon. However, within the range of parameter space probed by our simulations~(i.e.,
$10^{-5} \leq |\f0| \leq 10^{-4}$ for $\fr$ and $1089\,\mpc \leq r_c \leq 1665\,\mpc$ for Galileon), despite
having quite different cosmic growth histories, the two theories exhibit strikingly similar GIK deviations
from GR, with $\sim 100\text{--}200\,\kms$ enhancement in the characteristic infall velocity at $r=5\,\hmpc$,
and $\sim 50\text{--}100\,\kms$ broadening in the radial and tangential velocity dispersions across the infall
region, for clusters with mass $\sim 10^{14}\,\hmsol$ at $z=0.25$. These deviations are detectable through GIK
modelling of the redshift--space cluster--galaxy correlation function $\xicgs$, especially when combined with cluster
WL measurements. We highlight the imprint of MG on $\xicgs$ using the characteristic U-shaped curve of $\rpic$,
which increases by $\sim 1\text{--}2\,\hmpc$ at $r_p<6\,\hmpc$ from the GR prediction. We find little difference
between the GIK profiles predicted by the two screening mechanisms, except for slightly different trends of the
radial velocity dispersion with distance.

It is unclear whether the similar signature of these two distinct modified gravity theories on GIK is a coincidence,
or a generic result for any typical scalar--tensor theory that recovers the observed $\lcdm$--like expansion
history and reduces to normal gravity in the solar system and binary pulsars. In either case, our findings imply
that, in combination with WL, galaxy infall kinematics offer a powerful non--parametric cosmological test of
modified gravity. Ongoing galaxy redshift surveys will provide large samples of clusters with good statistics for
measuring $\xicgs$ and inferring GIK out to large scales. The main systematic uncertainty arises from the imperfect
understanding of the impact of galaxy formation physics on GIK. Within the context of GIK modelling and calibration,
the infall behavior of realistic galaxies could differ from that of tracers in cosmological simulations~(e.g.,
halos/sub--halos in N-body simulations, post--processed galaxies in semi--analytical galaxy formation models, and
simulated galaxies in hydrodynamic simulations, etc; See~\citeauthor{wu2013} 2013). However, we expect minimal
impact on the characteristic infall velocity, which is our main tool of estimating the dynamical mass profiles
of clusters, as any physical process that modifies galaxy kinematics within halos likely only adds scatter to the
velocity dispersions rather than changing the mean.

Within the context of testing gravity, the effects of galaxy formation physics could be partly degenerate with
those of modified gravity. The observed properties of galaxies, including luminosity, morphology, colour, star
formation, and clustering, are known to correlate with the environment~\citep[see, e.g., ][]{guo2013, zehavi2011,
zu2008, park2007, kauffmann2004, balogh2004, hogg2003, goto2003}, and MG theories also rely on the environment
to mediate the strength of the fifth force. For example, a sample of preferentially blue, star--forming galaxies
may show enhanced infall velocities compared to a galaxy sample that is unbiased in colour in $\lcdm$ universes,
and the enhancement could be mistaken as signal of modified gravity were the selection bias not properly accounted
for. We will investigate the potential systematic uncertainties induced by galaxy formation physics in a future
paper, using mock galaxy samples constructed from different HOD and semi-analytical model prescriptions. Redshift
surveys that probe a range of galaxy types are especially valuable  for those cosmological tests because one can check
that different classes of galaxies lead to the same cosmological conclusions even though the galaxy samples themselves
have different clustering and kinematics.

Our GIK modelling of $\xicgs$ is complementary to other semi--analytical approaches based on the halo
model~\citep{lam2013, lam2012}, both seeking to model the velocity distribution
around massive clusters for
testing gravity~\citep[also see][for an alternative method of modelling galaxy redshift--space distortion based on
HOD]{tinker2006, tinker2007}. The semi--analytical velocity model adopted in \cite{lam2013} has three components:
the empirical infall velocity from the spherical collapse model, the halo-halo pairwise velocity distribution, and
the intra-halo velocities (assumed Maxwellian with constant scatter). While the model itself is highly informative,
the accuracy is slightly lacking compared to simulation predictions. We instead use the simulations as emulators
for GIK, trading more computer time for better accuracy in the prediction of our model. In terms of observational
applications, our method differs from \cite{lam2013} in two significant aspects. First, they use the stacked
redshift differences as the observable, but the model predicts the LOS velocity dispersion, so they are affected
by the systematics in the subtraction of Hubble flow in the 2-halo term; in our method Hubble flow is naturally
incorporated in the calculation of $\xicgs$. Second, they consider the velocity distributions up to the second
moment, while we are able to model the entire $\p2d$ including all higher moments.

Established as one of the most powerful probes of dark energy, stacked WL analysis of clusters requires deep
imaging surveys that can simultaneously yield lensed background galaxies and foreground cluster sample.
Forecasts for Stage III and Stage IV dark energy experiments predict cluster WL constraints that
are competitive with supernovae, baryon acoustic oscillations, and cosmic shear~\citep[see][sections 6 and
8.4]{weinberg2013}. To complement WL as a cosmological test of gravity, GIK modelling of galaxy clusters requires
overlap with a large galaxy redshift survey, such as the ongoing Baryon Oscillation Spectroscopic
Survey~\citep[BOSS,][]{dawson2013}, its higher redshift successor
eBOSS~\citep[see][]{comparat2013}, and the deeper
surveys planned for future facilities such as BigBOSS~\citep{schlegel2009}, DESpec~\citep{abdalla2012}, the Subaru
Prime Focus Spectrograph~\citep{ellis2012}, {\it Euclid}~\citep{laureijs2011}, and {\it WFIRST}~\citep{green2012,
spergel2013}. We expect that, in combination with the stacked cluster WL analysis, the redshift--space
cluster--galaxy cross--correlations can reveal an accurate and complete picture of the average galaxy infall
around clusters, allowing stringent tests of modified gravity theories for the origin of the accelerating
expansion of the Universe.

\section*{Acknowledgements}

We thank Lam Hui and Bhuvnesh Jain for helpful discussions. Y.Z. acknowledges the hospitality of the Dept. of Physics
and Astronomy at the University of Pennsylvania where he enjoyed a fruitful discussion with participants of the ``Novel
Probes of Gravity and Dark Energy'' Workshop. D.H.W. and Y.Z. are supported by the NSF grant AST-1009505. Y.Z. is
also supported by the Ohio State University through the Distinguished University Fellowship.  B.L. is supported
by the Royal Astronomical Society and Durham University.  E.J. acknowledges the support of a grant from the Simons
Foundation, award number 184549. M.W. and E.J. were partially supported by, and some of the numerical simulations
reported here were performed on a cluster supported in part by, the Kavli Institute for Cosmological Physics at the
University of Chicago through grants through grants NSF PHY-0114422 and NSF PHY-0551142 and an endowment from the
Kavli Foundation and its founder Fred Kavli.  We also acknowledge resources provided by the University of Chicago
Research Computing Center. M.W. was additionally supported by U.S. Dept. of Energy contract DE-FG02-90ER-40560.

%%%%%%%%%%%%%%%%%%%%%%%%%%%%%%%%%%%%%%%%%%%%%%%%%%%%%%
%  Bibliography
%%%%%%%%%%%%%%%%%%%%%%%%%%%%%%%%%%%%%%%%%%%%%%%%%%%%%%%

% \nocite{*}

\footnotesize{
\bibliographystyle{mn2e}
% \bibliography{gik-mg}
% \begin{thebibliography}{74}
% \expandafter\ifx\csname natexlab\endcsname\relax\def\natexlab#1{#1}\fi

}

\end{document}